\documentclass[sigconf]{acmart}

\AtBeginDocument{%
  \providecommand\BibTeX{{%
    \normalfont B\kern-0.5em{\scshape i\kern-0.25em b}\kern-0.8em\TeX}}}

\setcopyright{acmcopyright}
\copyrightyear{2023}
\acmYear{2023}
\acmConference[AI4S]{the 4th workshop on Artificial Intelligence and Machine Learning for Scientific Applications, SC23}{November 12--17,
  2023}{Denver, CO}
\usepackage{cleveref}
\begin{document}

\title{Accelerating Particle and Fluid Simulations with Differentiable Graph Networks for Solving Forward and Inverse Problems}

\author{Krishna Kumar}
\email{krishnak@utexas.edu}
\orcid{0000-0003-2144-5562}
\affiliation{%
  \institution{University of Texas at Austin}
  \streetaddress{ECJ 9.227 B, 301 E. Dean Keaton St.}
  \city{Austin}
  \state{Texas}
  \country{USA}
  \postcode{78751}
}

\author{Yongjin Choi}
\email{yjchoi@utexas.edu}
\affiliation{%
  \institution{University of Texas at Austin}
  \streetaddress{ECJ 9.227 B, 301 E. Dean Keaton St.}
  \city{Austin}
  \state{Texas}
  \country{USA}
  \postcode{78751}
}

\renewcommand{\shortauthors}{Kumar and Choi}

\begin{abstract}
We leverage physics-embedded differentiable graph network simulators (GNS) to accelerate particulate and fluid simulations to solve forward and inverse problems. 
GNS represents the domain as a graph with particles as nodes and learned interactions as edges. 
Compared to modeling global dynamics, GNS enables learning local interaction laws through edge messages, improving its generalization to new environments. 
GNS achieves over 165x speedup for granular flow prediction compared to parallel CPU numerical simulations. 
We propose a novel hybrid GNS/Material Point Method (MPM) to accelerate forward simulations by minimizing error on a pure surrogate model by interleaving MPM in GNS rollouts to satisfy conservation laws  and minimize errors achieving 24x speedup compared to pure numerical simulations. 
The differentiable GNS enables solving inverse problems through automatic differentiation, identifying material parameters that result in target runout distances. 
We demonstrate the ability of GNS to solve inverse problems by iteratively updating the friction angle (a material property) by computing the gradient of a loss function based on the final and target runouts, thereby identifying the friction angle that best matches the observed runout.
The physics-embedded and differentiable simulators open an exciting new paradigm for AI-accelerated design, control, and optimization.
\end{abstract}




\begin{teaserfigure}
  \includegraphics[width=\textwidth]{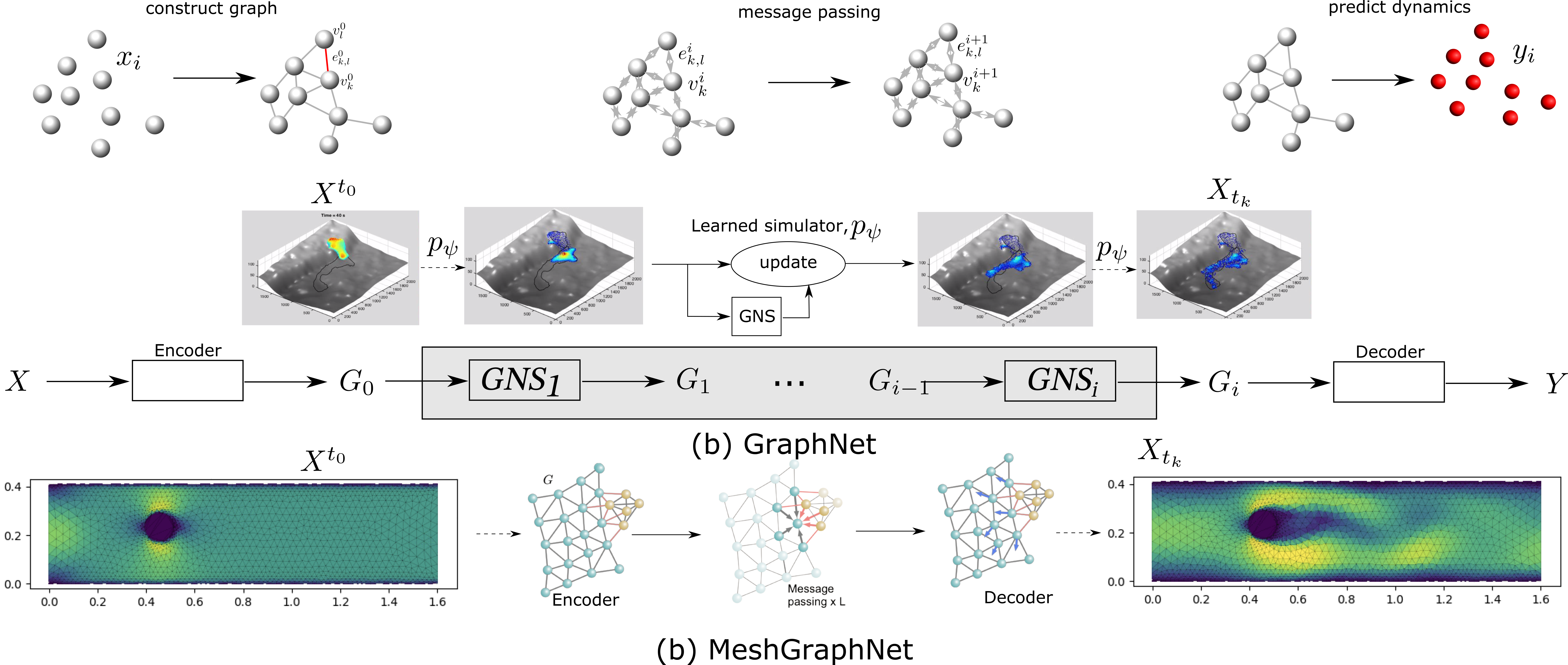}
  \caption{Graph network and MeshNet simulator for accelerating particulate and fluid simulations (modified after~\cite{sanchez2020learning})}
  \Description{Graph network simulator for accelerating particulate simulations}
  \label{fig:teaser}
\end{teaserfigure}

\received{11 August 2023}

\maketitle

\section{Introduction}
Simulators that realistically capture complex physics, such as particulate and fluid flow, provide immense value across scientific and engineering fields. Particulate systems such as granular media show complex transitionary behavior between solid-like and fluid-like responses. Additionally, the turbulent behavior of fluid flow poses unique challenges in modeling their flow around complex boundaries. These simulations require fine-mesh resolutions to capture intricate geometries and long compute times to converge on solutions. To make such simulations more practical approaches like reduced-order models are often used but sacrifice accuracy for efficiency. Conventional continuum-based simulation techniques, such as the finite element or finite difference methods, can model small-strain problems and face mesh distortion issues in modeling large-deformation flow problems~\cite{soga2016trends}. Although hybrid Eulerian-Lagrangian methods such as the Material Point Method (MPM) can simulate large deformation particulate flow problems, such as landslides, they are computationally expensive and are limited to representative elemental volume with at most 1M particles; in contrast, a cubic meter of soil has more than 1 billion grains. AI algorithms are widely adopted in building data-only surrogate models; however, they are often used as black boxes to predict a single outcome, such as failure or no failure, and lack physics~\cite{DoE2019}. 

We develop a physics-embedded graph network simulator (GNS) that represents the domain as nodes and interactions as learned edge functions, allowing it to generalize beyond training regimes. Using an attention-based GNS surrogate, we propose a novel physics-embedded framework for developing surrogate models. Furthermore, we accelerate the computational efficiency of traditional simulators while minimizing the errors in a data-only surrogate by proposing a novel hybrid GNS/MPM. The hybrid GNS/MPM combines the best of both works by interleaving MPM with GNS to conserve physics laws while accelerating the forward simulations with GNS, offering an order of magnitude better computational efficiency over a pure numerical simulation. A major challenge in studying particulate and fluid flow is solving optimization and inverse problems. This inverse analysis of identifying the optimal configuration or material properties that yield a specific response requires incrementally varying input parameters/model configuration and rerunning models to match observations - an inefficient trial-and-error approach. By exploiting automatic differentiation in GNS, we solve the inverse analysis with gradient-based optimization. Furthermore, The NextGen differentiable GNS opens a new AI-embedded design, control, and optimization paradigm.

\section{State of the art}
Numerical methods provide approximate solutions to partial differential equations (PDEs) by discretizing the solution space into finite entities. Particle-based approaches like the discrete element method (DEM) offer the advantage of modeling the microscale grain-grain interactions, albeit constrained to representative elemental volumes~\citep{kumar2019large}. Traditional continuum-based methodologies, such as the finite element method (FEM), are proficient in predicting failure initiation but fall short due to mesh distortions when handling large-deformation runouts~\citep{soga2016trends}. Hybrid Eulerian-Lagrangian methods like the material point method (MPM) alleviate mesh distortion issues but necessitate grid and material point tracking, proving computationally expensive~\citep{kumar2019scalable}. However, these methods only leverage CPU parallelization, and the hybrid particle-mesh transfer degrades the scaling performance of MPM, limiting its applicability for exa-scale simulations. Furthermore, traditional forward simulators cannot solve inverse and design problems, as they are limited to computing gradients in the forward mode. Solving inverse problems requires a special adjoint method that manually defines the derivative of the forward model equations. The lack of reverse-mode differentiation limits the AI-embedded simulation paradigm.

Neural network (NN)-based ML models have shown promising results in predicting soil deformations under specific load conditions~\citep{andrade2019multiscale,mital2021bridging,wang2019cooperative,wang2019deep}. However, these models' `black-box' nature impedes interpretability, necessitating significant training data and leaving them vulnerable to adversarial attacks. Physics-informed neural networks (PINNs) embed prior knowledge, such as PDEs and boundary conditions, as a loss function in model training ~\cite {raissi2019physics}. However, PINNs are limited to the boundary conditions of the training data and may not yield PDE-compliant predictions during extrapolation. Graph network simulators (GNS) offer a promising alternative that exploits graph networks to represent the underlying domain and learn the local interaction rather than the global dynamics, thus allowing extrapolation to geometries beyond the training regime~\citep{battaglia2018relational,sanchez2020learning,kumar2022gns}.~\citet{haeri2021accurate} reduced the dimensionality of the data using Principal Component Analysis to model graph networks.~\citet{mayr2023boundary} developed a contact boundary in GNS to model complex boundary interactions with granular media.~\citet{kumar2022gns} developed a multi-GPU parallel GNS to achieve linear strong scaling during GNS training.~\citet{kumar2022minority} exploited GNS as an oracle for large-scale in situ visualization of regional-scale landslides.

A new class of differentiable simulators offers a promising solution to solve complex inverse problems by enabling differentiation in forward and reverse modes through automatic differentiation. Initiatives like JAX-MD~\cite{schoenholz2020jax} and JAX-FLUIDS have made strides towards creating differentiable simulators (DiffSim) for particulate and fluid systems~\cite {hu2019difftaichi}. Differentiable simulation allows the incorporation of physics and domain knowledge into ML models, leading to better generalization. They enable end-to-end gradient-based optimization facilitating continuous adaptation and meta-learning. Nevertheless, the object-oriented design of numerical methods, replete with branching conditions, poses challenges in automatic differentiation, requiring stateless implementations for acceleration with Just-In-Time compilations. Integrating AI acceleration with traditional numerical methods can revolutionize numerical simulations and achieve new simulation frontiers.

\section{Graph Network Simulation}\label{sec:gns}
Graphs can represent interactions in physical systems \citep{battaglia2018relational, sanchez2020learning}. We represent the particulate media as a graph $G=(\boldsymbol{V}, \boldsymbol{E})$ consisting of a set of vertices ($\boldsymbol{v}_{i} \in \boldsymbol{V}$) representing the particles or aggregation of particles and edges ($\boldsymbol{e}_{i,j} \in \boldsymbol{E}$) connecting a pair of vertices ($\boldsymbol{v}_i$ and $\boldsymbol{v}_j$) representing the interaction between them. Graphs offer a permutation-invariant form of encoding data, where the interaction between vertices is independent of the order of vertices or their position in Euclidean space.

Graph neural network (GNN) takes a graph $G=(\boldsymbol{V},\boldsymbol{E})$ as an input, computes properties and updates the graph, and outputs an updated graph $G'=(\boldsymbol{V}', \boldsymbol{E}')$  with an identical structure, where $\boldsymbol{V}'$ and $\boldsymbol{E}'$ are the set of updated vertex and edge features ($\boldsymbol{v}_i'$ and $\boldsymbol{e}_{i, j}'$). GNN generates an updated graph by propagating information through the graph, termed \textit{message passing}.

Graph Network Simulators (GNS)~\citep{sanchez2020learning, kumar2022gns, kumar2022minority, choi2023graph} operate on graphs to learn the physics of the dynamic system and predict rollouts. The graph network spans the system domain with nodes representing a collection of particles and the links connecting the nodes representing the local interaction between particles or clusters of particles. The GNS learns the physics of the system dynamics, such as momentum and energy exchange, through message passing on the graph. GNS has three components (see~\cref{fig:teaser}a): (a) Encoder, which embeds particle information to a latent graph, the edges are learned functions; (b) Processor, which allows data propagation and computes the nodal interactions across steps; and (c) Decoder, which extracts the relevant dynamics (e.g., particle acceleration) from the graph. We introduce physics-inspired inductive biases, such as an inertial frame that allows learning algorithms to prioritize one solution (constant gravitational acceleration) over another, reducing learning time. The GNS implementation uses semi-implicit Euler integration to update the next state based on the predicted accelerations. We extend GNS with an attention mechanism to focus on the local interaction law and generate physically consistent predictions by enforcing conservation laws (mass, momentum, and energy) as soft constraints. The attention coefficient between nodes is defined as a weighted function of the feature over its neighbors. The graph attention mechanism improves predictions over long-time scales with weight-sharing properties to represent dynamically changing neighbors typical in large-deformation particulate flows. 

\subsection{Training and rollout}
The training datasets include 26 square-shaped granular mass flow trajectories in a two-dimensional box boundary simulated using the Material Point Method (CB-Geo MPM) code \citep{kumar2019scalable}. Each simulation has a different initial configuration regarding the size of the square granular mass, position, and velocity. We used a learning rate $\eta = 1E-4$ and trained for 20M epochs on Nvidia A100 GPU nodes on TACC LoneStar6. 

GNS successfully predicts the rollout of granular media within 5\% particle location error compared to MPM simulations (see~\cref{fig:hybrid-gns-mpm}). Additionally, GNS achieves a speed-up greater than 165x compared with distributed memory parallel CB-Geo MPM code. 

\subsection{MeshGraphNet}
\begin{figure}
    \centering
    \includegraphics[width=\linewidth]{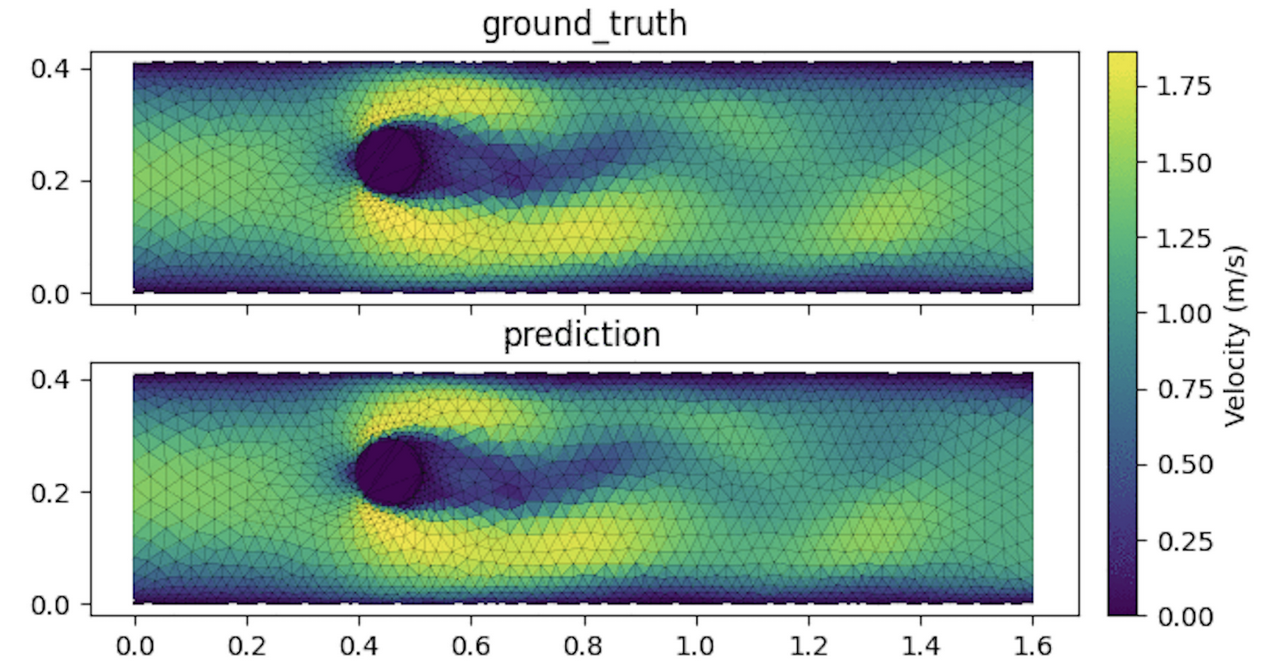}
    \caption{MeshNet for simulating fluid flow.}
    \label{fig:meshnet}
    \vspace{-1em}
\end{figure}

We describe the state of the system at time t using a simulation mesh $M_t = (V, E_M)$ with nodes
$V$ connected by mesh edges $E_M$ . Each node $i \in V$ is associated with a reference mesh-space
coordinate $x_i$, which spans the simulation mesh and additional dynamical quantities $q_i$ that we want to model. 
The task is to learn a forward model of the dynamic quantities of the mesh at time $t+1$ given the current mesh state $M_t$ and a history of previous meshes ${M_{t-1}, \dots, M_{t-n}}$.
We employ a similar architecture to the graph neural network of an Encode-Process-Decode architecture, followed by an integrator, as shown in~\cref{fig:teaser}b.~\Cref{fig:meshnet} shows the prediction of a von Karman vortex shedding from the MeshGraphNet compared with a ground truth Computational Fluid Dynamics (CFD) solution.

\section{Accelerating forward problems with GNS}
\begin{figure*}
    \centering
    \includegraphics[width=0.8\linewidth]{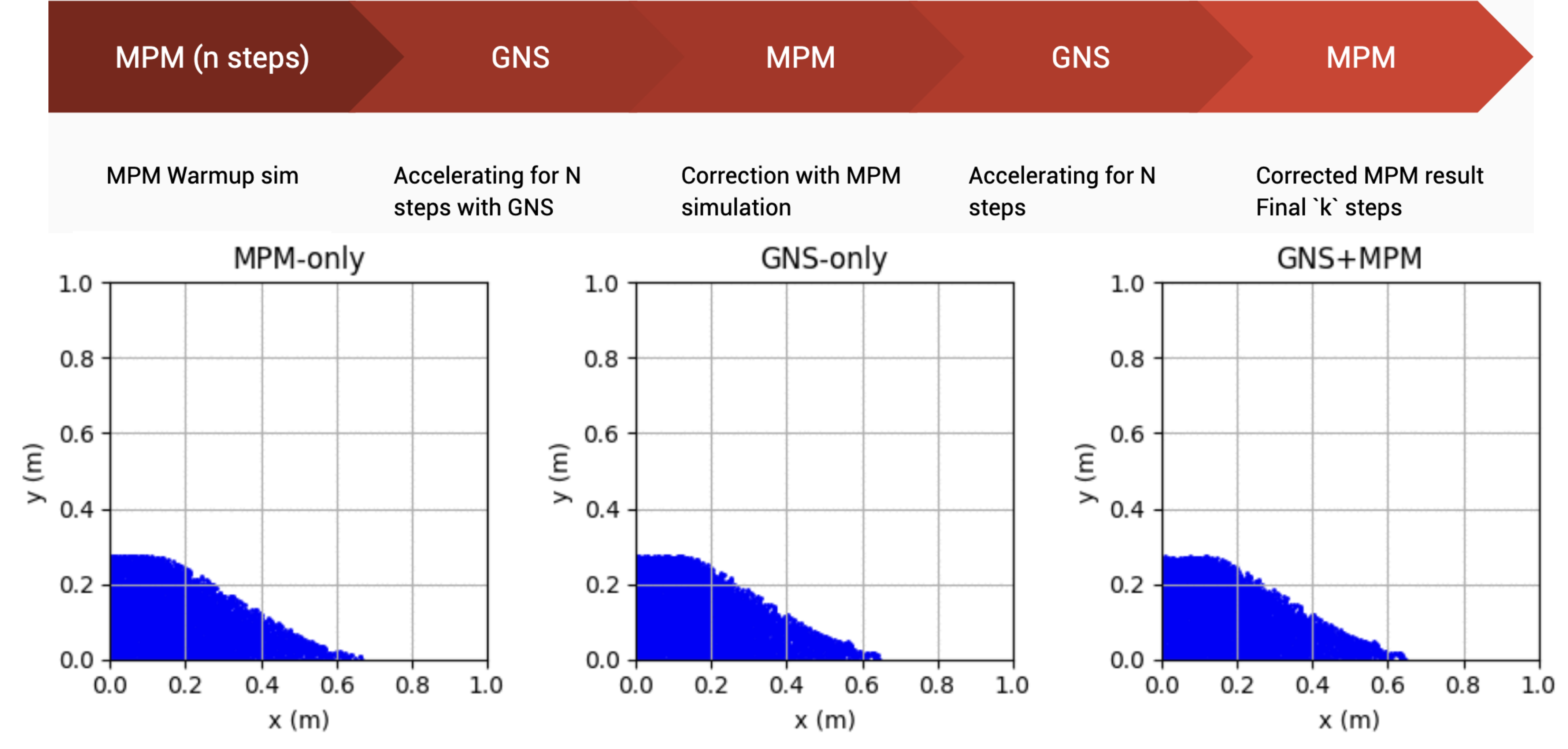}
    \caption{Accelerating forward simulation with hybrid GNS/MPM.}
    \label{fig:hybrid-gns-mpm}
\end{figure*}

\begin{figure}
    \centering
    \includegraphics[width=\linewidth]{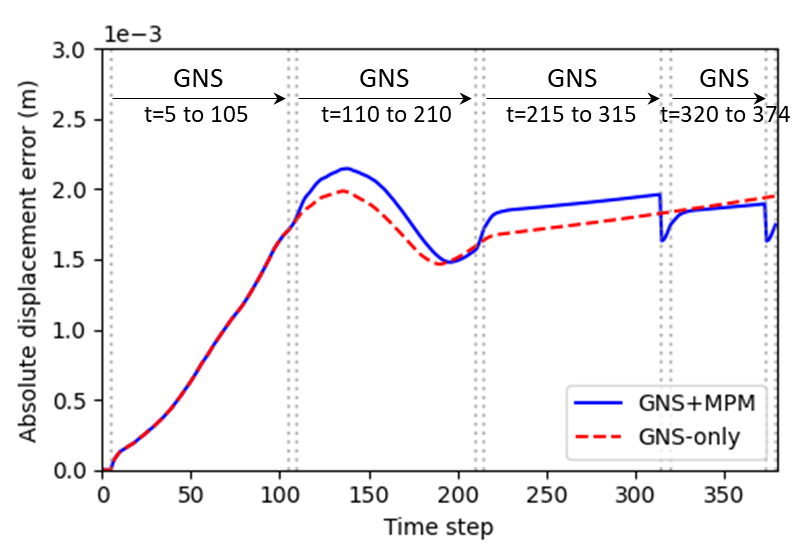}
    \caption{Hybrid GNS/MPM error evolution compared to GNS.}
    \label{fig:gns-mpm-eeror}
\end{figure}

We develop a hybrid GNS-accelerated numerical simulation with the Material Point Method for a fast solution to forward problems. We design a hybrid GNS-MPM approach incorporating domain-specific knowledge and conservation laws to achieve improved convergence.~\Cref{fig:hybrid-gns-mpm} shows the hybrid GNS-MPM framework, which includes three main stages. \textit{Warm-up}: GNS prediction requires the previous five steps to predict a rollout. We first generate the initial five velocity steps using MPM with specified boundary conditions. We run the physics solver with a predefined `$K$' of five steps. \textit{GNS rollout}: After the warm-up step, we predict the rollout, as described in~\cref{sec:gns}, based on the previous K timesteps for further `$M$' steps. \textit{Iterative Refinement}: The output of the GNS rollout may not satisfy known conservation laws, despite inductive biases and constraints. We feed the output of the GNN rollout to the MPM physics solver to perform `$K$' iterations. The data-driven model integrated with MPM will generate physics-conserving simulations in less time. We achieve a speed-up of 20x compared to traditional explicit simulation, while most of the computation time is still spent on the `$n*K$' runs.~\Cref{fig:hybrid-gns-mpm} shows that the hybrid GNS+MPM reduces displacement errors compared to pure GNS-only runs.~\Cref{fig:gns-mpm-eeror} shows the effect of hybrid GNS+MPM in reducing the final error in pure-GNS-only models. Further research could explore different criteria for adaptive-switching between GNS/MPM based on error metrics. 

\section{Accelerating Inverse problems with Differentiable GNS} 
A critical challenge in engineering design and optimization is solving the inverse problem, which involves identifying the parameters that lead to a desired result. Traditional simulators like MPM can differentiate in the forward mode to compute gradients of PDEs. However, they cannot compute gradients needed for inverse problems using reverse-mode differentiation. Inverse problems require techniques like the adjoint method to manually define derivatives of the forward model equations to calculate gradients in reverse mode. We leverage automatic differentiation (AD) in the PyTorch version of GNS to solve inverse problems. AD uses the chain rule to compute gradients of complex differentiable functions efficiently. AD enables accurate and fast gradient calculations by breaking down functions into elementary operations.

Our goal is to solve an inverse problem in granular flow to identify material properties that, given an initial geometry, result in a desired runout. We demonstrate this in the granular column collapse experiment. In this experiment, a rectangular granular column is released on a flat surface and collapses under gravity. The runout depends on the initial aspect ratio and material properties like friction angle. The inverse problem is to find the optimal friction angle ($\phi$) that gives a target runout distance ($L_f^{\phi_{target}}$) for a given initial column geometry (aspect ratio $a$). An optimizer computes the squared error between the target and simulated runout distances ( $J(L_f^{\phi_{target}}, L_f^{\phi})= (L_f^{\phi_{target}}- L_f^{\phi})^2$) and updates $\phi$ to minimize the error. We use AD to directly compute $\frac{\partial J}{\partial\phi}$ rather than finite differences.

The downside of using AD is that it requires a significant amount of memory for large-scale inversion of neural networks~\citep{allen2022inverse} because it retains all the gradients of parameters for all the intermediate layers during the backward pass. Since GNS contains multiple MLPs with multiple layers, and the entire simulation even entails the accumulation of positions $GNS (\boldsymbol{X}_t\rightarrow \boldsymbol{X}_{t+1})$ for $k$ steps, computing $\frac{\partial J}{\partial\phi}$ requires extensive memory capacity. We found that conducting AD for entire timesteps is not feasible in the currently available GPU memory capacity (40 GB). For this reason, we conduct the AD on the CPU and restrict the forward pass to $k$=30 steps in the optimization process. Accordingly, our target runout corresponds to the runout at 30 steps, not at the final timestep when the flow reaches static equilibrium.

\begin{figure}
    \centering
    \includegraphics[width=0.9\linewidth]{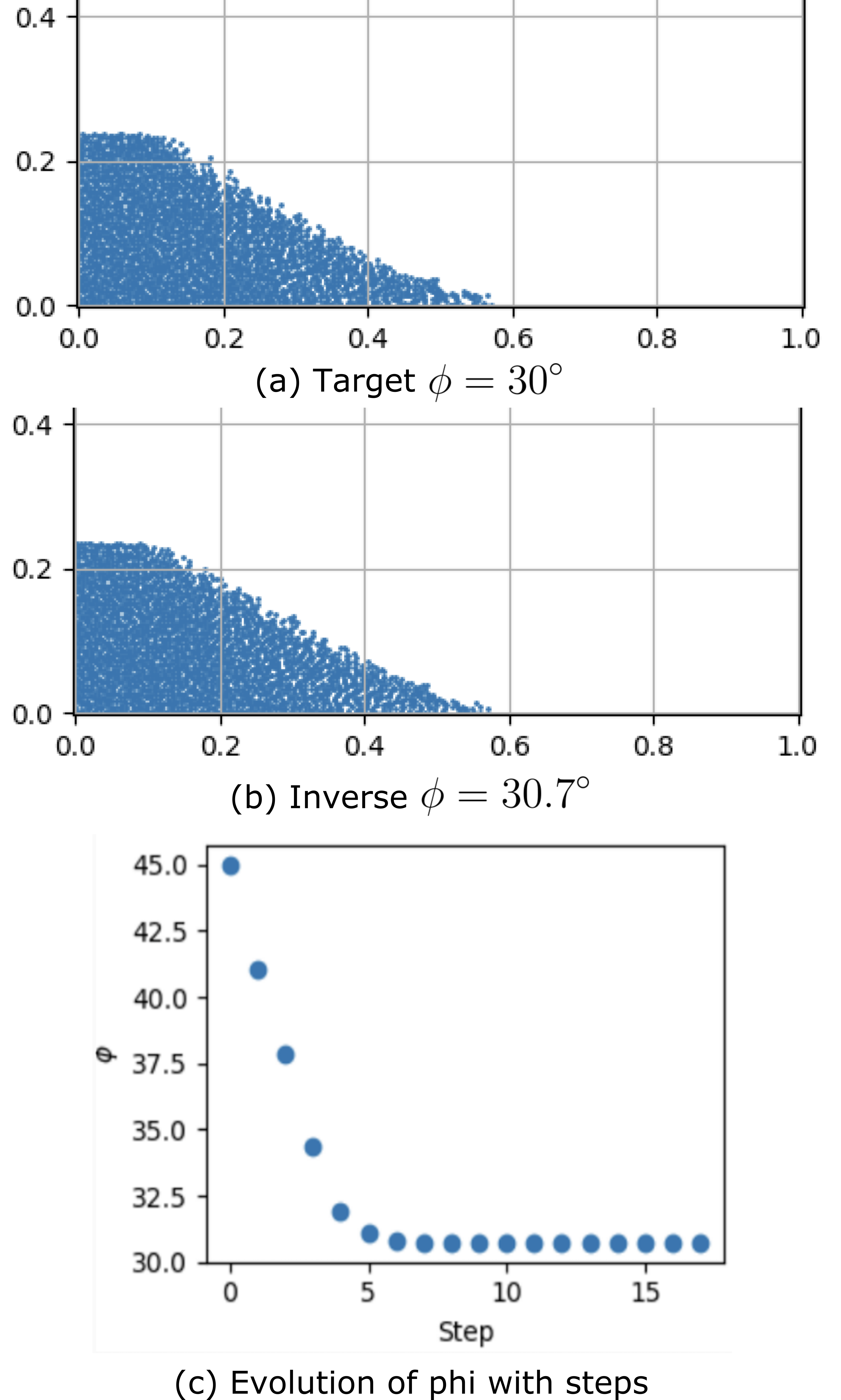}
    \caption{Solving inverse problems with GNS.}
    \label{fig:inverse}
\end{figure}

\Cref{fig:inverse}a shows the target profile for a friction angle $\phi = 30^\circ$. We use an initial guess of $\phi = 45^\circ$ to solve the inverse problem of estimating the friction angle based on the final runout profile. We use a simple gradient descent algorithm to update the friction angle at each step based on the gradient of the loss function with respect to the friction angle. After 17 iterations, the solution converges to $\phi = 30.7^\circ$ (see~\cref{fig:inverse}b).~\Cref{fig:inverse} shows the evolution of friction with each inverse iteration step. The friction angle converges quickly in about six iterations. We demonstrate that a single-parameter inversion based on the runout distance successfully identifies the initial material properties based only on the final runout by computing gradients using automatic differentiation.

\section{Interpretable GNS}
When a GNS successfully replicates a physics system's dynamics, we hypothesize that the messages encoding the latent information preserve the interaction laws. 
The sparse representation of the GNS messages ($e^\prime_k \leftarrow \phi^e(e_k, v_{r_k}, v_{s_k}, u)$) is a learned linear combination of the true forces.

\begin{figure}
    \centering
    \includegraphics[width=0.9\linewidth]{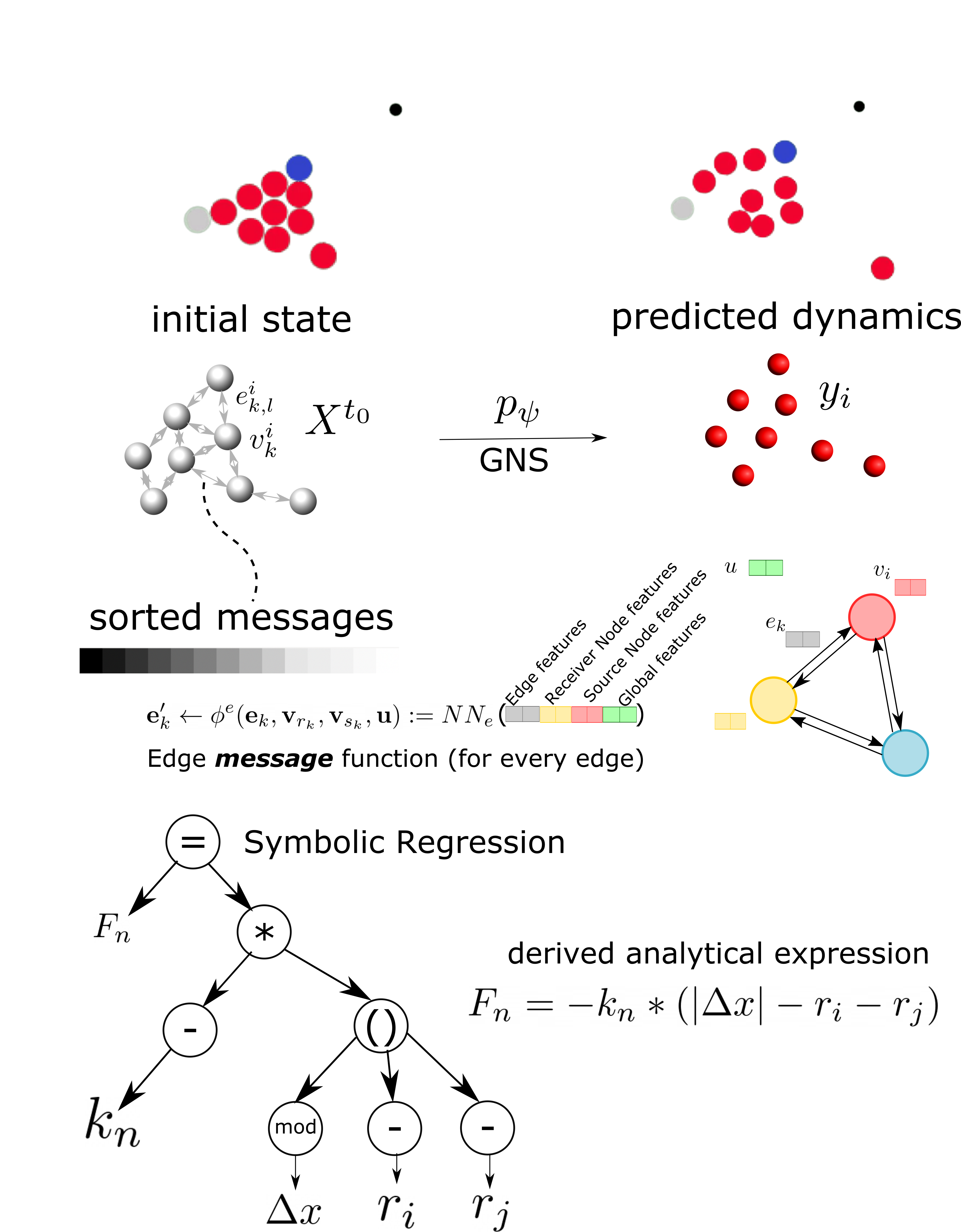}
    \caption{GNN simulation of N-body dynamics and Symbolic Regression explanation of edge interaction.}
    \label{fig:gnn-sr}
\end{figure}

We predict the n-body dynamics using the open-source data-parallel PyTorch GNS code developed by the PI~\cite{kumar2022gns,kumar2022insitu}. 
The GNN is trained on 30 different trajectories of n-body dynamics ($\sim 10$ particles) for 1 million steps. For particle $i$, $m_i$ is the mass, $k_{n_i}$ is its stiffness, $r_i$ is the radius, $x_i$ is the position. The message $e^\prime_k \leftarrow \phi^e(e_k, v_{r_k}, v_{s_k}, u)$ contains the edge features ($e_k = f(k_{n}, \gamma_{n}, \delta_{n}$)), source and receiver vertex features $v_{s_k}, v_{r_k} = g(m_i, r_i, x_i)$ and no global features. We fit the most significant features by enforcing a sparsity constraint on the messages through L1 regularization, which forces us to learn the minimal vector space required to describe the messages. Furthermore, we restrict the number of message components by sorting them based on the largest standard deviation. 

We then take 10,000 randomly selected outputs from our testing set to derive the force law from the output of vertex and edge neural networks. We derive the physics laws by approximating the message data from the testing set with symbolic regression. Symbolic regression fits a function $\psi$ using the following edge and vertex features ($m_i, m_j, r_i, r_j, x_i, x_j, k_{ij}, \gamma_{ij})$ by minimizing the mean absolute error (MAE) through brute force genetic algorithm. The symbolic regression considers the following operators $+$, $-$, $*$, $/$, $>$, $<$, $pow$, $exp$, $inv$, $\log$ as well as real constants in its solutions. This task uses a simple algorithm to quantify the complexity $c$ by counting the number of occurrences of each operator, constant and variable. We weigh $pow, exp, inv, \log$ as three times the other operators to consider the complexity $C_x$ of the operation. We use an approach analogous to Occam’s razor to find the “best” algebraic model that minimizes errors at different complexity levels. We used a simple weighted counting model to quantify the complexity of the expression. We identify the symbolic expression as the one that maximizes the fractional drop in MAE over an increase in complexity from the next best model ($- \Delta \log (MAE_c)/\Delta c$). 

In this work, we extract the GNS edge messages of a small-scale system (10 bodies) interacting via linear springs. 
We then apply SR on GNS messages to identify the most accurate closed-form expression that describes the encoded interaction law as shown in~\cref{table:sr}. 
SR on GNS messages successfully derived (Eq 8 in~\cref{table:sr}) the force interaction law $F_n = k_n * abs(\Delta x - r_i - r_j)$ of a linear spring with stiffness $k_n = 100$, relative position $\Delta x$  between two particles ($i$ and $j$) and their radii $r$. 

\begin{table*}[]
\caption{Symbolic regression derivation of force interaction law from observing GNN messages of 10-body collision of linear-spring system.}
\label{table:sr}
\begin{tabular}{l l l c c}
\toprule
\textbf{Eq.} & \textbf{Derived equation}                                  & \textbf{MSE}       & \textbf{$C_x$} & \textbf{$D_a$}\\
\midrule
1 & $-198.72363$                                            & 64747.52  & 1          & Y                   \\
2 & $(\Delta x + -198.92792)$                               & 63938.996 & 3          & N                   \\
3 & $(-203.1408 + exp(\Delta x))$                           & 60907.68  & 4          & N                   \\
4 & $((\Delta x + -2.3484528) * 92.79602)$                  & 27031.744 & 5          & Y                   \\
5 & $((\Delta x * (\mathbf{X}_1 + 92.75565)) + -218.16481)$ & 26830.58  & 7          & N                   \\
6 & $((\Delta x + (abs(r_1) * -1.1491286)) * 100.23312)$    & 21227.312 & 8          & Y                   \\
7 & $((\Delta x + ((abs(r_1) + 1.1013538) * -0.8038518)) * 98.86028)$                  & 18721.219         & 10          & Y          \\
\textbf{8*} & $\mathbf{((\Delta x + (abs((r_2 * -1.0) + r_1) * -1.0)) * 100.0)}$                 & \textbf{3.76E-10} & \textbf{12} & \textbf{Y} \\
9 & $((\Delta x + (abs((r_2 * inv(-1.0)) + r_1) * -1.0)) * 99.9998)$ & 3.01E-10          & 15          & N    \\     
\bottomrule
\multicolumn{4}{l}{$C_x$ is the complexity and $D_a$ represents if the expression passes dimensional analysis.}\\
\multicolumn{4}{l}{$*$ denotes the chosen solution.}\\
\end{tabular}
\end{table*}

\section{Limitations}
While the graph neural network simulator demonstrates promising acceleration for simple particulate systems, applying it to diverse large-scale multi-physics problems poses significant research challenges. The current node-level attention mechanism needs further analysis on its ability to learn interaction physics and generalize across problems effectively. Scaling GNS using graph partitioning and advanced sampling techniques are essential for training GNS on millions of particles. Furthermore, orchestrating hybrid GNS/MPM framework using accurate error metrics to determine when to switch between data-driven prediction and physical solvers is an important direction. Overcoming these limitations in generalization, scalability, physical fidelity, and hybrid modeling will be vital to unlocking the potential of differentiable GNS for accelerating scientific discoveries.

\section{Conclusions}
This work introduces novel physics-embedded differentiable graph network simulators (GNS) to accelerate particle and fluid simulations and solve challenging inverse problems. The graph representation allows learning localized physics interactions compared to global dynamics, improving generalization. GNS achieves over 165x speedup compared to parallel CPU MPM simulations for granular flow prediction. The differentiable GNS enables solving inverse problems through automatic differentiation, identifying material parameters that result in target runout distances. The physics-embedded and differentiable simulators open an exciting new paradigm for AI-accelerated design, control, and optimization.

\section{Acknowledgments}
This material is based upon work supported by the National Science Foundation under Grant No.\#2103937. Any opinions, findings, and conclusions or recommendations expressed in this material are those of the author(s) and do not necessarily reflect the views of the National Science Foundation.

\bibliographystyle{ACM-Reference-Format}
\bibliography{paper}

% \appendix

\end{document}